# Confronting the Carbon-footprint Challenge of Blockchain

Xiaoyang Shi,[1,‡,*] Hang Xiao,[2,‡] Weifeng Liu,[3,4] Xi Chen[1], Klaus. S. Lackner[5], Vitalik Buterin[6],

Thomas F. Stocker[7]

[1] Earth Engineering Center, Center for Advanced Materials for Energy and Environment, Department of Earth and Environmental Engineering, Columbia University, New York, NY 10027, USA

[2] Shanxi Institute of Energy and Chemical Engineering, School of Chemical Engineering, Northwest University, Xi'an 710069, China

[3] College of Hydrology and Water Resources, Hohai University, Xikang Road, Nanjing 210098, China

[4] State Key Laboratory of Simulation and Regulation of Water Cycle in River Basin, China Institute of Water Resources and Hydropower Research, Beijing 100038, China

[5] School of Sustainable Engineering & Built Environment, Arizona State University, Tempe, AZ 85287-9309, USA

[6] Ethereum Foundation

[7] Climate and Environmental Physics and Oeschger Center for Climate Change Research, University of Bern, CH-3012 Bern, Switzerland

[‡] These authors contributed equally.

Corresponding Author: xs2144@columbia.edu

**Abstract:** The distributed consensus mechanism is the backbone of the rapidly developing blockchain network. Blockchain platforms consume vast amounts of electricity based on the current consensus mechanism of Proof-of-Work (PoW). Here, we point out an advanced consensus mechanism named Proof-of-Stake (PoS) that can eliminate the extensive energy consumption of the current PoW-based blockchain. We comprehensively elucidate the current and projected energy consumption and carbon footprint of the PoW and PoS-based Bitcoin and Ethereum blockchain platforms. The model of energy consumption of PoS-based Ethereum blockchain can lead the way towards the prediction of other PoS-based blockchain technologies in the future. With the widespread adoption of blockchain technology, if the current PoW mechanism continues to be employed, the carbon footprint of Bitcoin and Ethereum can push the global temperature above 1.5 °C in this century. However, a PoS-based blockchain can reduce the carbon footprint by 99% compared to the PoW mechanism. The small amount of carbon footprint from PoS-based blockchain could make blockchain an attractive technology in a climate friendly future. The study sheds light on the urgency and effectiveness of developing the PoS mechanism to solve the current sustainability problem of blockchain.

## Main:

Over the last few years, blockchain technology has been proclaimed by many as the most significant technological breakthrough with the potential to transform most industries[1], such as finance, health care, Internet-of-Things networks, video games, and others[2,3]. With the help of cryptographic hash functions, digital signatures, and distributed consensus mechanisms, the record in the distributed database can remove the middlemen and establish trust between unknown parties[4]. For example, blockchain-based smart contracts can be partially or fully executed or enforced without human interaction when certain conditions are met. Blockchain-based digital currencies, another famous blockchain application allowing payment accessible to anyone in the world, had a total global market capitalization over $2.5 trillion at its peak.

For all its benefits and forward-facing applications, blockchain uses the energy-intensive method of the Proof-of-Work (PoW) consensus mechanism to verify and validate transactions. The validation of ownerships and transactions is based on network participants adding valid blocks to the chain through solving puzzles involving hash functions[5]. Adding a block requires network participants to deploy computational resources to evaluate hash functions that requires a certain amount of work. Because many active computational nodes are solving the same hash simultaneously, total energy consumption is large, resulting in a large carbon footprint that has an enormous impact on the environment[6]. Mora et al.[7] claim Bitcoin emissions alone could push global warming above 2°C within less than three decades. Masanet et al.[8] have pointed out that the scenarios used by Mora et al. are fundamentally flawed, and provided their three corrected scenarios that show the Bitcoin could push global warming exceeds 2°C by 2100.

The accumulation of carbon dioxide ($CO_2$) in the atmosphere is the primary driver of global warming[9]. Human forcing of climate change is increasing the probability of severe ecological impacts[10] and possibly the crossing of tipping points[11,12]. According to scenario RCP 8.5 reported in Intergovernmental Panel on Climate Change (IPCC), without intervention, $CO_2$ emissions will rise from the current 49 $GtCO_{2eq}$/yr to between 85 and 136 $GtCO_{2eq}$/yr by 2050[13]. The rising $CO_2$ concentration could cause a global mean temperature change from the pre-industrial level (1880-1900) to 2100 of 3.8 to 6.0 °C. The global annual carbon emission from Bitcoin and Ethereum mining is 43.9 Megaton $CO_2$ ($MtCO_2$) in 2017[14]. The energy consumption of blockchain is expected to grow by several orders of magnitude in the future[8]. Such growth has been observed based on the experience with other technologies, e.g., in the case of debit cards, stoves, and automobiles.

Proposed solutions to the carbon-intensive problem in blockchain include developing efficient hardware, reducing the difficulty required to resolve puzzles, or utilizing renewable energy. Each method has its own limitations; eg., the saved energy from the more efficient hardware may be employed to solve more puzzles for additional reward. The additional reward further incentivizes the purchase of more hardware to increase mining hashrate and the chances of higher rewards. This situation may also cause other miners to follow the trend to purchase more hardware. The vicious circle also limits the motivation for the innovation of more efficient equipment; reducing the puzzle difficulty may attract more computing power onto the network; the production of renewable energy is subject to seasonality and may not meet the consistent power demand of blockchain. Here, we evaluate the current energy consumption and carbon footprint of PoW-based blockchain platforms, and estimate the projected carbon footprint based on a new consensus mechanism, i.e., Proof-of-Stake (PoS). The PoS mechanism eliminates solving puzzles and thus dramatically reduces electric power consumption. Instead, in the PoS mechanism, the participants are selected by an algorithm for the right to validate blocks. With PoS, the probability of validating a block depends on the amount of stake a blockchain participant holds. The results of our model of energy consumption and carbon footprint from PoS-based blockchain demonstrate that PoS mechanism can reduce carbon emissions by two orders of magnitude compared to PoW mechanism. The migration from PoW to PoS may fundamentally solve the urgent energy-intensity problem in the blockchain system.

**Proof-of-Work Consensus Mechanism**

PoW-based blockchain, like Bitcoin, uses enormous amounts of energy to secure its network. Validation of blockchain by the Proof-of-Work consensus is computationally intensive. The PoW mechanism was first introduced in 1993 to combat spam emails[15] and was formally called "Proof-of-Work" in 1999[16]. This technology was not widely used until Satoshi Nakamoto created Bitcoin in 2009[4]. This mechanism could be employed to reach consensus between many nodes on a network and thereby secure the Bitcoin blockchain. However, the PoW mechanism works by employing many nodes to solve a cryptographic puzzle. The miner who first finds the solution receives a mining reward in the form of Bitcoin. The PoW consensus process is shown in Fig. 1a. based on previous studies[5,17,18]. The main drivers of the energy consumption of Bitcoin were found to be the geographical distribution of miners and the efficiency of the mining equipment[19]. de Vries proposed a market dynamics approach to evaluate the current methods for obtaining the Bitcoin consumes 87.1 TWh of electrical energy annually per September 30, 2019[20]. The energy consumption of Bitcoin is enough to power 9 million households in the US or the entire country of Belgium or Chile[21]. de Vries[22] also claimed the total network of Bitcoin could consume up to 184 TWh per year with 60% of miner's income going to pay for electricity at a price of $0.05 per kWh if the price of BTC is $42,000. For comparison, the current total annual electricity consumption is about 26,000 TWh in the world.

**The Migration from Proof-of-Work to Proof-of-Stake Consensus Mechanism**

Eliminating the high energy consumption of blockchain is critical to its survival. Aiming to overcome the issues of energy consumption and low transaction speed, moving to the PoS mechanism is an effective path for the sustainable development of Bitcoin and other PoW-based blockchain platforms. There are various difficulties in migrating the PoW mechanism to the PoS mechanism on Bitcoin, such as a conflict of interest of miners. However, Bitcoin can be built on the PoS mechanism in theory; a potential way to do this could be to leverage the pre-existing full and light node network to implement a PoS style solution. Once the PoS mechanism on other blockchain platforms is proven, that Bitcoin can adapt to it.[23]

Ethereum[24] pioneers a plan for transition from the current energy-intensive and low-throughput PoW mechanism to a more robust, efficient and energy-saving PoS mechanism. The upgrade of Ethereum from 1.0 (PoW mechanism) to 2.0 (PoS mechanism) includes three phases: Phase 0 - Beacon Chain, Phase 1 - Shard Chains, and Phase 2 - State Execution[25]. The mainnet of Phase 0 – Beacon Chain has been launched in Dec. 2020, and more than 140000 validator clients have joined the mainnet. The migration process of Ethereum has been introduced in the Supplementary Materials. The success migration on Ethereum can be used as a reference for Bitcoin. Therefore, this study analyzes the architecture of the PoS mechanism on Ethereum and calculates the energy consumption and carbon footprint of both PoW and PoS mechanisms. This study will provide the basis for Ethereum to deliver an accurate carbon footprint of its PoS mechanism. The mathematical model of energy consumption of PoS-based Ethereum blockchain can lead the way towards the prediction of energy consumptions of other PoS-based blockchain technologies in the future.

**Proof-of-Stake Consensus Mechanism**

The first Proof-of-Stake cryptocurrency, Peercoin[26], was developed by Sunny King and Scott Nadal in 2012. The stake in the Peercoin network is coin age, i.e., network tokens times holding period. However, participants in the Peercoin network still need to solve a cryptographic puzzle, whose difficulty decreases with a higher coin age. Thus, the Proof-of-Stake consensus of Peercoin is a hybrid PoS/PoW. In the state-of-the-art PoS networks, the coin age is replaced by the number of network tokens a participant holds, and the PoW puzzle is completely removed. Miners are replaced by stakers who lock up network tokens as stakes in the ecosystem. The proposers of the new block (who can receive the block reward) are selected based on their stakes in PoS networks, instead of their computational power in PoW networks. Fig. 1b shows the working process of the PoS mechanism.

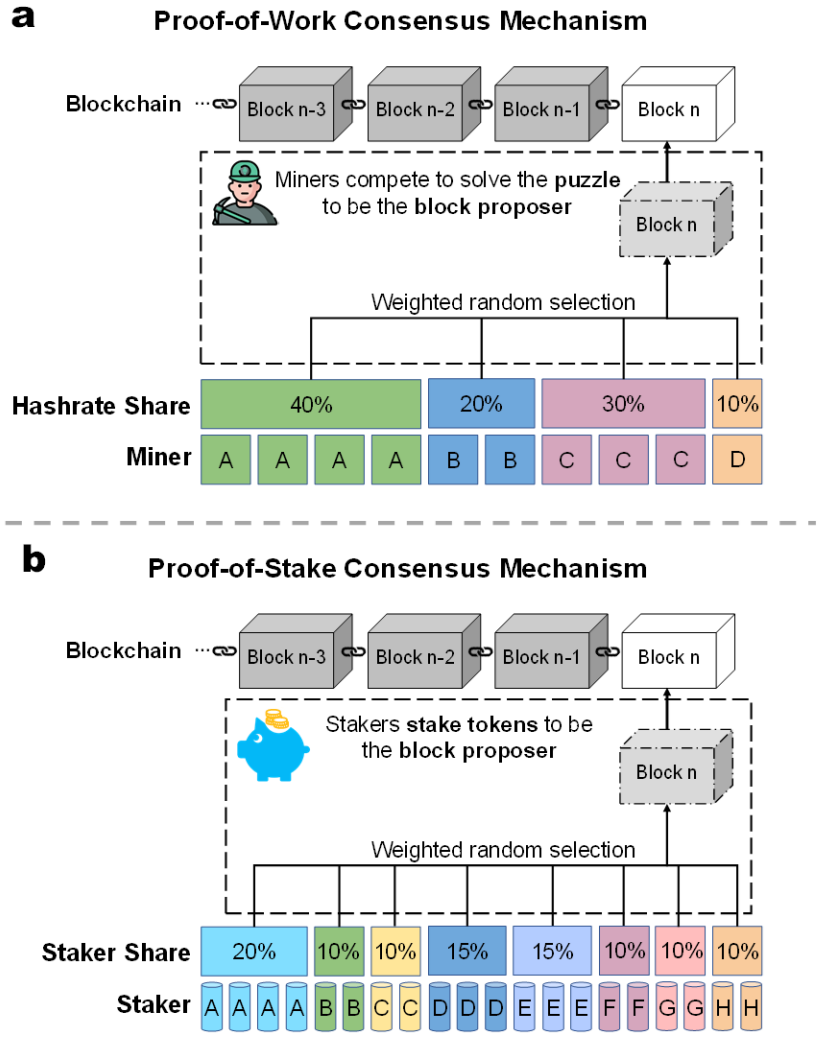

**Fig. 1.** Comparison of (**a**) Proof-of-Work and (**b**) Proof-of-Stake mechanisms. With Proof-of-work (**a**), miners own computing power to compete against each other to complete transactions on the network and get rewarded. The hashrate share represents the share of the computing power owned by one miner occupying the total computing power of the whole network. If a miner manages to solve the puzzle, the new "Block n" is considered confirmed and connected to the previous block. Quantum computer is able to break the mathematical difficulty underlying most of currently used cryptography. The only solution in this case is to transit to a new type of cryptography, which is considered to be inherently resistant to quantum attacks[27]. Proof-of-Stake (**b**) is a proposed alternative to PoW. With Proof-of-Stake, the resource that is compared is the amount of share a staker holds. For example, someone holding 1% of the Bitcoin (share) can own 1% of the probability to validate the next block. Different PoS-based platforms have some differences in their algorithms. For example, PoS of Ethereum is the underlying mechanism that activates
7777

validators upon receipt of enough stake. Users will need to stake 32 ETH (Ethereum Tokens) to become a validator. Validators are chosen randomly to create blocks and are responsible for checking and confirming blocks they do not create. A user's stake is also used as a way to incentivize good validator behavior[28]. Another example is Algorand that was launched in 2019 by Dr. Silvio Micali[29] and his team from MIT. Algorand uses a unique variation of Proof of Stake called Pure Proof of Stake (PPoS). PPoS is a highly democratized PoS consensus mechanism with a low minimum staking requirement for participating in and securing the network — only 1 ALGO coin is required to participate[29]. A PoS-based blockchain has more validators means a more secure and decentralized system.

**The Advantages and Disadvantages of Proof-of-Stake Mechanism**

The Proof-of-Stake mechanism has advantages over Proof-of-Work mechanism:

1. PoS is fairer compared to PoW. A staker must deposit an amount of coins into the network as a stake in PoS. The size of the stake determines the chances of a staker to be chosen to forge the next block which is a linear correlation. With PoW, rich participants can take advantage of the power of economies at scale. The more mining equipment they buy, the lower prices of per mining equipment they can obtain.

2. PoS is less expensive because PoS-based blockchain does not require costly mining equipment and extensive electricity for mining, but only requires hosting a node. This change encourages more participants to set up a node. The barriers to entry are lower and should remain at a constant level.

3. PoS makes the network more decentralized and secure. Staking is more decentralized thanks to the lower barrier to entry. PoW has mining pools where participants are teaming up to mine. These pools now control 80% mining power of the bitcoin blockchain. If a group of miners can obtain 51% of the hashing power, they can effectively manage the blockchain and start approving fraudulent transactions.

4. PoS is a more energy-efficient mechanism. PoS does not allow everyone to mine for new blocks by solving puzzles simultaneously and therefore uses considerably less energy. Validators on PoS are chosen at random to create blocks and are responsible for checking and confirming blocks they do not create, that allows PoS to handle higher scalability and larger throughput. The transaction verification time on PoW can be reduced by reducing its difficulty, or reducing the data size of transactions. These ways may increase the transaction throughput to a limited extent. However, the possibility of forks and vicious attacks cannot be ignored. PoS intends to provide scalability that is orders of magnitude greater than what is currently available PoW. For example, a jump from roughly 10 TPS on PoW-based Ethereum to potentially 100,000 TPS on PoS-based Ethereum.

PoS also has its concerns:

1. PoS has to carefully select the next block proposers. Developers need to design a fair algorithm. A fair algorithm factors in the size of the stake and progressively disfavors the holders of large numbers of stakes, i.e., adopt an "under-proportional" policy, select proposer = constant × log (stakes). Additionally, sharding design ensures that someone staking more tokens needs to validate more blocks without more rewards.
2. PoS has a weak subjectivity issue in which a client needs to log on at least a few times a year, or else validators who has already withdrew their deposits can attack that client at no cost. This issue is impossible to avoid entirely, but features like the withdrawal queue, allows the period the client can safely go between logging in without making validator withdrawals take too long in the normal case.
3. PoS is more complex to implement compared to PoW. There are many risks during the implementation of PoS, and serious flaws of code may be fatal. This is the main trade-off of PoS. PoW entrusts complexity to the power consumption of mining, so it can enhance security by using external resources. The security of PoS has to be solved by extremely careful designs and might still fail eventually. The threat of a 51% attack still exists in proof-of-stake, but it is riskier for the attackers. For example, if Bitcoin is running on PoS, attackers need to control 51% of the staked BTC execute an attack. Not only is the staked BTC worth hundreds of billions of dollars, but it would probably cause BTC's value to drop. There is very little incentive to destroy the value of a currency attackers have a majority stake in.

**The Energy Consumption and Carbon Footprint of PoW-based Ethereum and Bitcoin**

Here, we present a techno-economic model for determining the electricity consumption and carbon footprint of PoW based Ethereum and Bitcoin. Our analysis is based on mining revenue and hardware data for estimating energy consumption and carbon emission. Determining an accurate level of electricity consumption by Ethereum is unrealistic because of the great variety of Graphics Processing Units (GPUs) used for mining[30]. Instead, we bracket the solution range with an upper and a lower limit. The upper limit is set as the break-even point of mining revenues and electricity costs. Rational behavior would lead miners to disconnect their hardware from the network as soon as their costs exceed their revenues from mining and validation. For example, the price drop of ETH token drove miners to stop mining in 2019, reducing the upper limit of electricity consumption on Ethereum. The lower limit is defined by a scenario in which all miners use the most efficient hardware in each year[31], as reported in Data Sheet 1.1. The efficiency of the hardware is an essential parameter in determining energy consumption. The calculation details are introduced in the Supplementary Materials. Based on the geographical footprint of Ethereum nodes and the household electricity prices of the countries where nodes are located, we can calculate the weighted average of the electricity price and carbon emission factor of nodes running on the PoW-based Ethereum. The calculation is expressed in detail in Supplementary Materials and data is listed in Data Sheet 2.2-2.3. Fig. 2 shows the location of the nodes that are identified based on device IPs. The weighted average of the electricity price and carbon emission factor of nodes running on Bitcoin has been reported in Data Sheet 2.4-2.5.

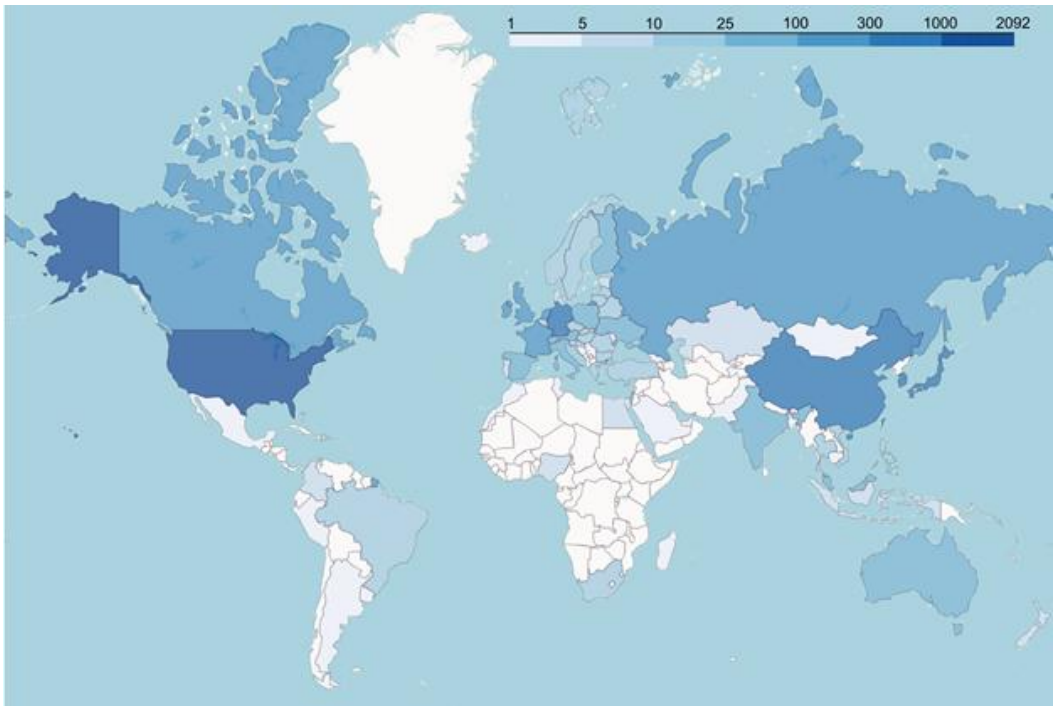

**Fig. 2.** The geographical footprint of Ethereum 1.0 full nodes as reported in Data Sheet 2.1. This geographic footprint allows for an accurate estimation of energy consumption and carbon emissions as reported in Data Sheet 2.6.

Fig. 3 shows annual electricity consumption and carbon footprint of Ethereum and Bitcoin. The figure shows that the annual electricity consumption/carbon footprint of the upper limit of Ethereum has increased from 0.67 TWh/0.32 $MtCO_2$ in 2016 to 11.91 TWh/5.15 $MtCO_2$ in 2020. The electricity consumption/carbon footprint of the lower limit follows the same trend as the upper limit, increasing from 0.27 TWh/0.13 $MtCO_2$ to 2.22 TWh/0.96 $MtCO_2$ from 2016 to 2020. The upper limit of Bitcoin has increased from 5.60 TWh/4.04 MtCO$_2$ in 2016 to 56.42 TWh/38.61 MtCO$_2$ in 2020. The lower limit of Bitcoin has increased from 1.31 TWh/0.63 MtCO$_2$ in 2016 to 31.50 TWh/13.61 MtCO$_2$ in 2020. Although the decrease in market price resulted in a reduction of electricity consumption in 2019 because of higher price decreases in the crypto market, the one order of magnitude increase of energy consumption since 2016 has raised concerns about the sustainability of blockchain technology due to its rapidly growing carbon footprint.

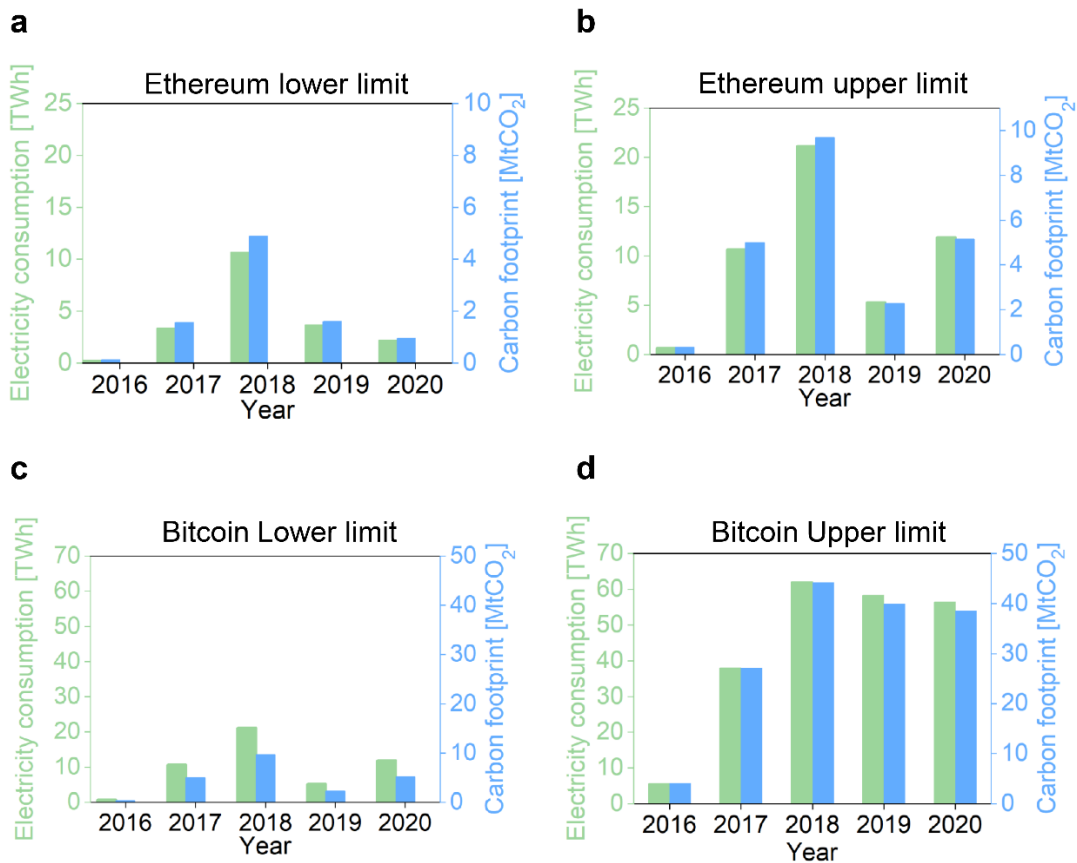

**Fig. 3.** Annual electricity consumption and carbon footprint estimates of the PoW-based Ethereum on its lower limit **(a)** and upper limit **(b),** and Bitcoin on its lower limit **(c)** and upper limit **(d)**. Note that the weighted average emission factor electricity price, internet price, and energy intensity of each country and each year from 2016-2010 were used to calculate the electricity and carbon emission of Ethereum and Bitcoin blockchain. These numbers are reported in Data: Sheet 2.2 – 2.6.

**The Energy Consumption and Carbon Footprint of PoS-based Ethereum**

A pseudo agent-based model of a blockchain system with PoS has been used to calculate the energy consumption[32]. Here, we build a realistic model based on Ethereum's PoS mechanism.

A typical staker setup on Ethereum 2.0 consists of one Beacon node program and multiple validator client programs running on a computer with Internet connection. Each validator client program requires a minimum stake of 32 ETH. On Ethereum 2.0, the computing cost of running the Beacon node program is identical to that of a full node program (e.g., Geth) on Ethereum 1.0[33]. The validator client program on Ethereum 2.0 plays a similar role as the miner on Ethereum 1.0. However, unlike miners on PoW-based Ethereum 1.0, the validator client program is resource-efficient, and its energy consumption is significantly less than that of the Beacon node program. As a result, the energy consumption of running validator client programs is assumed to be negligible.

The lower limit of the carbon footprint of Ethereum 2.0 is defined by a scenario in which all stakers use the energy-efficient Jetson TX2 module (5W) to run Beacon node programs and validator client programs. This lower limit of annual energy consumption and its carbon footprint for Ethereum 2.0 comes out at 0.0396 TWh and 0.0171 $MtCO_2$, respectively. In this scenario, the calculated total number of Beacon nodes running on Ethereum 2.0 is 903569 (See the calculation details in Supplementary Materials).

The upper limit of the carbon footprint of Ethereum 2.0 is defined by a scenario in which all stakers use the Intel Xeon server[34] to run Beacon node programs and validator client programs resulting in 0.3119 TWh and 0.1348 $MtCO_2$, respectively. In this scenario, the total number of Beacon nodes running on Ethereum 2.0 is 439507. Given that the total number of full nodes on Ethereum 1.0 is less than 7000, the PoS-based Ethereum 2.0 is expected to be capable of achieving a significantly higher level of decentralization with a much lower carbon footprint than PoW-based Ethereum 1.0.

**Projected Carbon Footprints of PoW and PoS-based Blockchain Platforms**

The future usage of blockchain is a topic of considerable discussion because it is a promising and revolutionary technology. Wide usage of blockchain would result in a large carbon footprint. The energy consumption of Ethereum and Bitcoin exceeds 88% of the energy consumption of the entire blockchain space[14]; therefore, we mainly consider Ethereum and Bitcoin in this section.

Blockchain usage has experienced accelerated growth (Supplementary Fig. 1), which is a typical pattern during the early adoption of broadly used technologies. Fig. 4a shows trends in the adoption of broadly used technologies over time. To develop a scenario for blockchain we used the incorporation rate of 36 different techniques for which data are readily available, such as debit cards, stoves, automobiles, water closets, and so on (See details in Data Sheet: 3.1-3.2). The inset figure in Fig. 4a shows the historical adoption rate of ETH and BTC in the past 5 and 11 years, respectively. If PoW-based Ethereum were to follow the median growth trend observed in the adoption of these 36 technologies, the cumulative $CO_2$ emissions of this blockchain alone could warm the planet 0.26 to 0.43 °C by 2120, shown in Fig. 4b. We conduct an uncertainty analysis on the impact of carbon emissions on global temperature. The correlation coefficient between carbon emissions and global temperature is from the temperature projections of 42 Earth system models developed for Coupled Model Intercomparison Project Phase 5 (CMIP5). Purple right Y axis shows lower bound of uncertainty by carbon emission on temperature change, orange right Y axis shows the average value on temperature change, and green right Y axis shows the upper bound on temperature change. Fig. 4d shows the total cumulative $CO_2$ emissions of PoW-based Ethereum and PoW-based Bitcoin could warm the planet 6.23 to 10.16 °C by 2120. Another logistic growth model and its accuracy to predict the carbon footprint of blockchain have been proposed in the Supplementary Materials. The development of Ethereum and Bitcoin is highly dynamic and sensitive. The two projection models existing reasonable assumptions provide a baseline of Life Cycle Analysis to show PoW-based blockchain consumes extensive energy nowadays and its energy cost will be more outrageous in the near future.

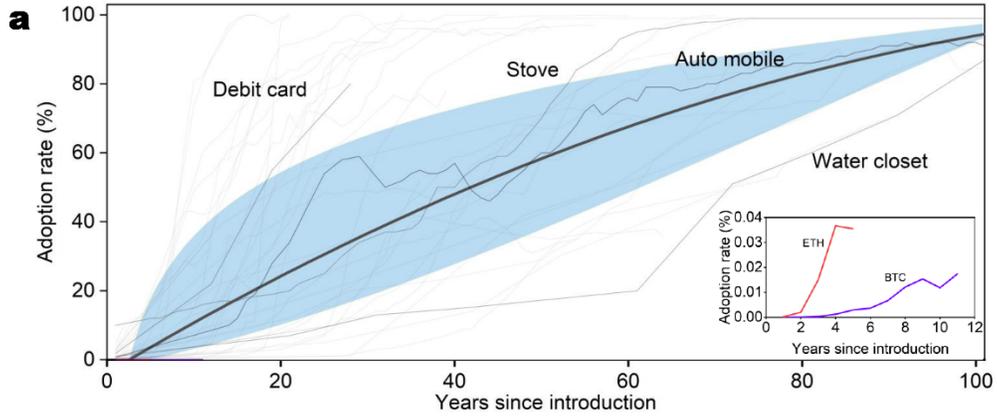

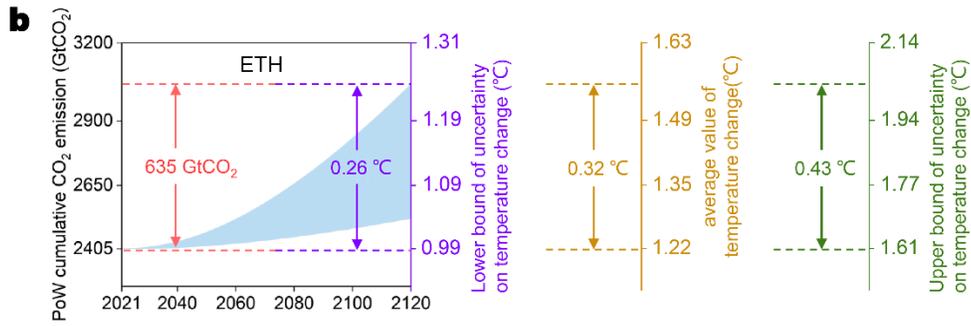

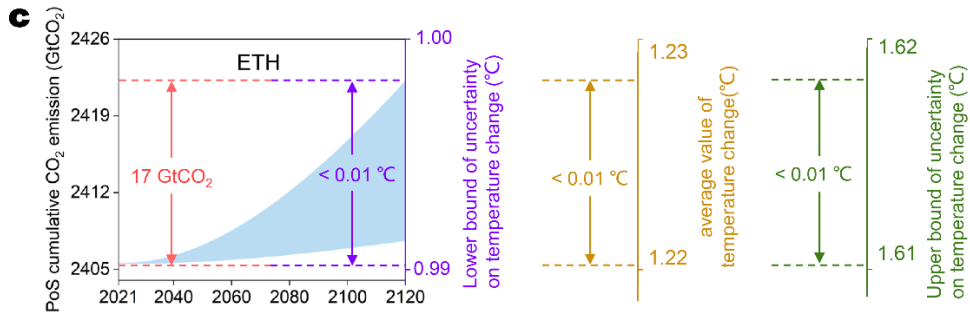

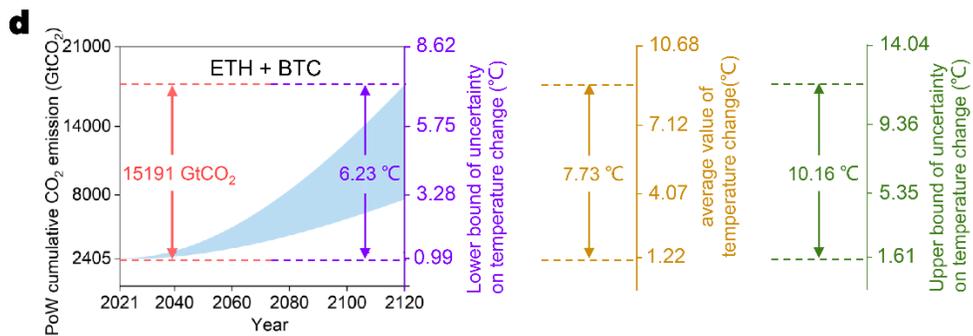

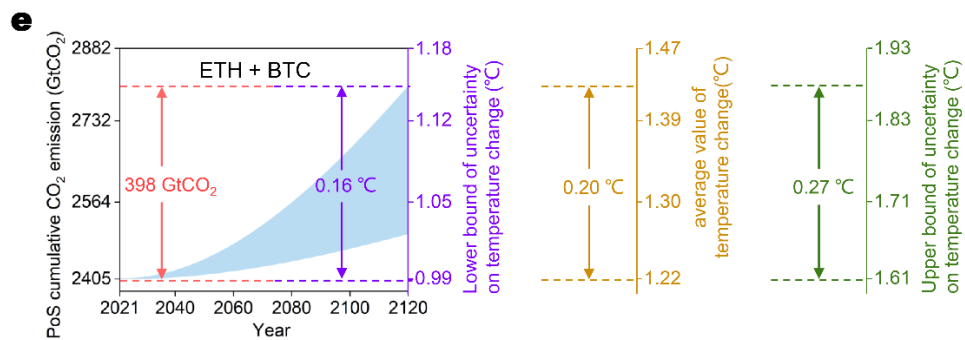

**Fig. 4** Projected Carbon Footprints of PoW and PoS-based Blockchain based on the adoption model. **Fig 4a** shows trends in the adoption of broadly used technologies documented in Data: Sheet 3.1. The blue shaded area indicates the margins of the upper 75% and lower 25% quantiles, and the thick black line is the median tendency across technologies (see Data: Sheet 3.1). Grey lines indicate trends for each of the technologies (see Data: Sheet 3.2). The inset figure shows the adoption rate of ETH and BTC of its own development in the past 5 and 11 years, respectively. **Fig 4b and 4c** show the projected cumulative $CO_2$ emissions and the resulting temperature rise from PoW-based Ethereum 1.0 and PoS-based Ethereum 2.0, respectively, if they were to follow the average growth rate of broadly adopted technologies. **Fig 4d** and **4e** show the projected cumulative $CO_2$ emissions and the resulting temperature rise from PoW-based Ethereum 1.0 plus PoW-based Bitcoin, and PoS-based Ethereum 2.0 plus PoS-based Bitcoin, respectively, under the average growth rate of technologies shown in **Fig. 4a**. We conduct an uncertainty analysis on the relationship between carbon emissions and the increase in global temperature. Purple right Y axis shows lower bound of uncertainty by carbon emission on temperature change, orange right Y axis shows the average value on temperature change, and green right Y axis shows the upper bound on temperature change. The Data is shown in Data: Sheets 3.1–3.8.

The utilization of renewable energy can reduce carbon emissions to some extent. One estimate claims 39% energy consumption of Bitcoin mining is from renewable energy[35]. However, this will not solve the sustainability problem of blockchain[36], which is not limited to Ethereum and Bitcoin but shared by the whole blockchain industry. Given the decentralized nature of blockchain, its computing verification process of the overall system must migrate to the PoS mechanism that could help reduce the carbon footprint of Ethereum by 99%. Fig. 4c shows in 100 years, the cumulative carbon emission from PoS-based Ethereum is 17 $GtCO_2$ pushing the temperature to increase less than 3% of the carbon emission from PoW-based Ethereum shown in Fig. 4b. We can estimate the energy consumption of PoS-based Bitcoin in accordance with the mathematical model of PoS-based Ethereum established in the study. Fig. 4d shows the carbon emission from PoS-based Ethereum and Bitcoin. The temperature increase from PoS-based Ethereum and Bitcoin can be reduced to 0.16 °C to 0.27 °C compared to 6.23 °C to 10.16 °C caused by PoW-based Ethereum and Bitcoin in 100 years at the case of the average value of temperature change. The energy consumption of different PoS-based blockchain platforms, such as Algorand[29] or Avalanche[37], are different due to their altered algorithms, but the energy consumption of all PoS-based blockchain platforms can reduce significantly in comparison to PoW-based blockchain. There are other potential ways could be deployed on blockchain for sustainable outcomes. For example, Delegated proof of stake (DPoS) and the 'delegated Byzantine fault tolerance' (DBFT) models use negligible amounts of energy to grant validating power to stake-holding nodes[38].

The study suggests that if the adoption rate of PoW-based blockchain follows other broadly used technologies or logistic growth model, it could create an electricity demand capable of producing enough $CO_2$ emissions to exceed 1.5 °C of global warming in decades. The analysis of electricity consumption and carbon emission elucidate that any further development of blockchain should shift towards to PoS mechanism to reduce the electricity demand if the consequences of 1.5 °C of global warming is aiming to be avoided. The change of the migration from PoW to PoS mechanism will profoundly affect the energy consumption and the sustainable development of blockchain.

**SUPPLEMENTAL INFORMATION**

Supplemental Information can be found online

**ACKNOWLEDGMENTS**

This work is supported from the Earth Engineering Center and Center for Advanced Materials for Energy and Environment, Columbia University. Klaus S. Lackner acknowledges the support by ASU Fulton Schools of Engineering Dean's Office. Weifeng Liu acknowledges the fellowship support from China Scholarship Council during his visit at Columbia University.

**Author Contributions**

T.F.S., K.S.L., and V.B. supervised the project. X.Y.S. conceived the idea. H.X. designed and carried out the model of PoS mechanism. X.Y.S. and W.F.L. designed and carried out the model of PoW mechanism. X.Y.S., W.F.L., and K.S.L. designed and performed the models of projected carbon footprint. X.Y.S., X.H., and W.F.L. analyzed the data. V.B. contributed to the analysis of PoW and PoS architecture of blockchain. K.S.L. and T.F.S. analyzed the impact of carbon footprint and energy consumption from PoW and PoS-based blockchain. X.Y.S., K.S.L., T.F.S., V.B., and H.X. wrote the manuscript. All authors contributed to the results discussion and manuscript preparation.

**DECLARATION OF INTERESTS**

All authors declare no competing interests.

# References:

# References:

1  Marsal-Llacuna, M.-L. Future living framework: Is blockchain the next enabling network? *Technological Forecasting and Social Change* **128**, 226-234, doi:https://doi.org/10.1016/j.techfore.2017.12.005 (2018).
2  Nguyen, C. T. *et al.* Proof-of-stake consensus mechanisms for future blockchain networks: fundamentals, applications and opportunities. *IEEE Access* **7**, 85727-85745 (2019).
3  Wang, W. *et al.* A Survey on Consensus Mechanisms and Mining Strategy Management in Blockchain Networks. *IEEE Access* **7**, 22328-22370, doi:10.1109/ACCESS.2019.2896108 (2019).
4  Nakamoto, S. Bitcoin: A peer-to-peer electronic cash system. https://bitcoin.org/bitcoin.pdf. (2008).
5  Narayanan, A., Bonneau, J., Felten, E., Miller, A. & Goldfeder, S. *Bitcoin and cryptocurrency technologies: a comprehensive introduction*. (Princeton University Press, 2016).
6  de Vries, A. Bitcoin's Growing Energy Problem. *Joule* **2**, 801-805, doi:https://doi.org/10.1016/j.joule.2018.04.016 (2018).

# Supplementary Information

Xiaoyang Shi,[1,‡,*] Hang Xiao,[2,‡] Weifeng Liu,[3,4] Xi Chen[1], Klaus. S. Lackner[5], Vitalik Buterin[6], Thomas F. Stocker[7]

[1] Earth Engineering Center, Center for Advanced Materials for Energy and Environment, Department of Earth and Environmental Engineering, Columbia University, New York, NY 10027, USA

[2] Shaanxi Institute of Energy and Chemical Engineering, School of Chemical Engineering, Northwest University, Xi'an 710069, China

[3] College of Hydrology and Water Resources, Hohai University, Xikang Road, Nanjing 210098, China

[4] State Key Laboratory of Simulation and Regulation of Water Cycle in River Basin, China Institute of Water Resources and Hydropower Research, Beijing 100038, China

[5] School of Sustainable Engineering & Built Environment, Arizona State University, Tempe, AZ 85287-9309, USA

[6] Ethereum Foundation

[7] Climate and Environmental Physics and Oeschger Center for Climate Change Research, University of Bern, CH-3012 Bern, Switzerland

[‡] These authors contributed equally.

Corresponding Author: xs2144@columbia.edu

**The Calculation of Energy Consumptions and Carbon Footprints of PoW Based ETH**

The lower limit is defined by a scenario in which all miners use the most efficient hardware in each year. We calculate the lower limit of the range by multiplying the required computing power—indicated by the hash rate—by the energy efficiency of the most efficient hardware:

Lower Limit

$$E_{Lower}^{PoW} = H * e_{ef} * 10^{-3} * 24 \tag{1}$$

with

$E_{Lower}^{PoW}$ = power consumption of lower limit [MWh]

$H$ = hash rate [GH/s]

$e_{ef}$ = energy efficiency of most efficient hardware [J/MH]

The upper limit is defined by the break-even point of revenues and electricity cost. Rational behavior would lead miners to disconnect their hardware from the network as soon as their costs exceed their revenues from mining and validation:

Upper Limit

$$E_{Upper}^{PoW} = \frac{(R_B + R_T + R_U + R_{UI}) * M}{p_N} * \frac{1}{1000} \tag{2}$$

with

$E_{Upper}^{PoW}$ = average daily power consumption of upper limit [MWh]

$R_B$ = block reward [ETH]

$R_T$ = transaction fees [ETH]

$R_U$ = uncle reward [ETH]

$R_{UI}$ = uncle including reward [ETH]

$M$ = market price [USD/ETH]

$p_N$ = electricity price [USD/kWh]

Based on the geographical footprint of Ethereum nodes and the household electricity prices of the countries where nodes are located, we can calculate the weighted average of the electricity price and carbon emission factor of nodes running on the Ethereum 1.0 network. The weighted carbon factor *n* is 0.4592 kg $CO_2$eq/kWh. The lower limit and upper limit of carbon footprint is $E_{Lower}^{Carbon} = E_{Lower}^{PoW} * n$ and $E_{Upper}^{Carbon} = E_{Upper}^{PoW} * n$, respectively.

**The Migration of Ethereum from PoW Consensus Mechanism to PoS Consensus Mechanism**

The upgrade of Ethereum from 1.0 (PoW consensus mechanism) to 2.0 (PoS consensus mechanism) includes 3 phases: Phase 0 - Beacon Chain, Phase 1 - Shard Chains, and Phase 2 - State Execution[1].

Phase 0 will launch the PoS Beacon Chain. It is the most crucial step of the Ethereum 2.0 upgrade since it introduces Casper[2], a PoS consensus mechanism to the network for the first time. The Beacon Chain is the core system-level blockchain of Ethereum 2.0, and it manages the PoS protocol for itself in Phase 0. In Phase 1, the Beacon chain will also manage all of the shard chains. Specifically, the functions of Beacon Chain will include (1) managing stakers and their stakes; (2)

randomly selecting block proposers based on stakes; (3) applying rewards and penalties to stakers; (4) facilitating cross-shard transactions (in phase 1 and onwards).

In Phase 1, the introduction of shard chains will further increase transactions throughput of Ethereum 2.0 by dividing the network across multiple shards, achieving parallel processing of transactions. Sometime after phase 1, the existing Ethereum 1.0 PoW chain will be shut down, and its accounts, states, as well as its applications will be moved to the Ethereum 2.0 PoS chain.

In Phase 2, Ethereum flavored WebAssembly (eWASM) that supports PoS and sharding will replace the Ethereum Virtual Machine (EVM). Smart contracts and accounts will be migrated from the PoW-based Ethereum 1.0 network to PoS-based Ethereum 2.0 network. After phase 2, Ethereum team plans to work on optimizations using cryptography primarily, for example polynomial commitments, snarks, to make the blockchain safer and cheaper to verify.

On Ethereum 1.0, a full node is a computer connected to the Ethereum network that stores the entire blockchain and enforces the consensus rules of Ethereum. Miners on Ethereum 1.0 need to connect to a full node to validate the new blocks by solving the PoW puzzle. The energy consumption of Ethereum 1.0 is primarily attributed to the process of solving PoW puzzles. Loghin et al[3] reported that the energy-efficient Jetson TX2 module can run an Ethereum full node with 5W power consumption, which could serve as the lower bound for the energy consumption of an Ethereum full node. The power consumption of a high-performance Intel Xeon server running an Ethereum full node is 81W[3], which could serve as the upper bound for the energy consumption of an Ethereum full node.

The mainnet of Phase 0 – Beacon Chain has been launched in Dec. 2020[4], and more than 140000 validator clients have joined the mainnet. Note that the specifications of the Shard Chains and eWASM are not finalized and are still under extensive research and development. Meanwhile, the Beacon Chain, which manages all the Shard Chains and applies the consensus rules, accounts for the dominant part of the energy consumption of the Ethereum 2.0 network. In the following, we predict the carbon footprint of Ethereum 2.0 network based on the specifications of the Beacon Chain.

**The Calculation of Energy Consumptions and Carbon Footprints of PoS Based ETH**

Note that staking more than 32 ETH on a validator client program does not generate more reward than staking exactly 32 ETH. Thus, we assume that all the validator client programs on Ethereum 2.0 network stake 32 ETH. So the total number of validator client programs on Ethereum 2.0, $N_{val}$, is

$$N_{val} = ETH_{stake}/32 \qquad (3)$$

where $ETH_{stake}$ is the total ETH stake on Ethereum 2.0. Assuming zero downtime for all validator client programs, the annual return of running a single validator client program, $g_{val}$, is[5]

$$g_{val} = \frac{5792.6176 * p_{ETH}}{\sqrt{ETH_{stake}}} \qquad (4)$$

where $p_{ETH}$ is the average price of ETH in a year (in US dollar per ETH [USD/ETH]). In the following, we will use the average price of ETH in 2020 to calculate the projected carbon footprint of Ethereum 2.0.

$$p_{ETH} = 307.5429 \text{ [USD/ETH]} \qquad (5)$$

Based on the geographical footprint of nodes and the household electricity prices of the countries where nodes are located, we can calculate the weighted average of the electricity price of nodes running on Ethereum 2.0 network, $\bar{p}_{ele}$ (see Data Sheet 2.3: Electricity_Internet):

$$\bar{p}_{ele} = 0.1783 \text{ [USD/kWh]} \tag{6}$$

Similarly, we can obtain the weighted average of the internet price of nodes running on Ethereum 2.0 network, $\bar{p}_{int}$ (see Data Sheet 2.3: Electricity_Internet):

$$\bar{p}_{int} = 39.5777 \text{ [USD/month]} \tag{7}$$

The weighted average of the emission factor of nodes running on Ethereum 2.0 network, $\bar{e}_{ef}$ (see Data sheet 2.2: Emission factors, we apply emission factors from the IEA[6]):

$$\bar{e}_{ef} = 0.4323 \text{ [kgCO2eq/kWh]} \tag{8}$$

The staker cost consists of computer cost, electricity cost and Internet cost. The depreciation period for computer is commonly 3 years. Thus, the expression of annual cost of a staker on Ethereum 2.0, $c_{stake}$, is

$$c_{stake} = \frac{p_{com}}{3} + \frac{365 * 24 * E_{com} * \bar{p}_{ele}}{1000} + 12 * \bar{p}_{int} \tag{9}$$

with

$p_{com}$ = computer price [USD]

$E_{com}$ = energy consumption of a computer running a Beacon node program [W]

We assume that all stakers use the typical staker setup mentioned above, and the ratio of the validator client programs to Beacon node programs running on a computer, $r_{val}$, is defined by the break-even point of staking rewards and staker cost. The expression of $r_{val}$ is

$$r_{val} = c_{stake}/g_{val} \tag{10}$$

Because it is assumed that all validators adopt the typical setup mentioned above. The total number of Beacon nodes, $N_{node}$, is

$$N_{node} = N_{val}/r_{val} \tag{11}$$

Combining equation (3,4,10,11), we derive:

$$N_{node} = \frac{\sqrt{ETH_{stake}/32} * 5792.6176 * p_{ETH}}{p_{com}/3 + 365*24/1000 * E_{com} * \bar{p}_{ele} + 12 * \bar{p}_{int}} \tag{12}$$

The projected annual energy consumption of Ethereum 2.0 network, $E$ [TWh], is

$$E = \frac{365 * 24 * E_{com} * N_{node}}{10^{12}} \tag{13}$$

Combining equation (12-13), we derive:

$$E = \frac{365 * 24 * E_{com} * \sqrt{ETH_{stake}/32} * 5792.6176 * p_{ETH}}{10^{12} * (p_{com}/3 + 365*24/1000 * E_{com} * \bar{p}_{ele} + 12 * \bar{p}_{int})} \tag{14}$$

The projected annual carbon footprint of Ethereum 2.0 network, $C$ [MtCO$_2$], is

$$C = E * \bar{e}_{ef} \tag{15}$$

Combining equation (14,15), we derive:

$$C = \frac{365 * 24 * E_{com} * \sqrt{ETH_{stake}/32} * 5792.6176 * p_{ETH} * \bar{e}_{ef}}{10^{12} * (p_{com}/3 + 365*24/1000 * E_{com} * \bar{p}_{ele} + 12 * \bar{p}_{int})} \tag{16}$$

According to equation (16), $C$ increases as $ETH_{stake}$ increases. On March 10$^{th}$, 2020, the total supply of Ethereum is 110030966 ETH. In the following discussion, we assume:

$$ETH_{stake} = 110030966 \text{ [ETH]} \qquad (17)$$

That is, we assume that 100% ETH on Ethereum network is staked.

The lower limit of the carbon footprint of Ethereum 2.0 is defined by a scenario in which all stakers use the energy-efficient Jetson TX2 module to run Beacon node programs and validator client programs. The price of Jetson TX2 module is 490.64 USD[7], so $p_{com} = 490.64$ USD. As mentioned above, Jetson TX2 is expected to run a Beacon node program at $E_{com} = 5$ W. By substituting $p_{com}$ and $E_{com}$ of Jetson TX2 and equation (5,6,7,8,17) into equation (14) and (16), we obtain the lower limit of energy consumption and carbon footprint of Ethereum 2.0 are 0.0396 TWh and 0.0171 $MtCO_2$, respectively. In this scenario, the total number of Beacon nodes running on Ethereum 2.0 is 903569.

The upper limit of the carbon footprint of Ethereum 2.0 is defined by a scenario in which all stakers use the high-performance Intel Xeon server[3] to run Beacon node programs and validator client programs. The Dell hardware customization service is utilized to calculate the price of the Intel Xeon server. The CPU of Intel Xeon server is Intel® Xeon® E5-1650 v3, which is discontinued. We use the currently available Intel® Xeon® E-2246G (of similar performance with Intel® Xeon® E5-1650 v3) to replace the discontinued Intel® Xeon® E5-1650 v3 in evaluating the price of the server. The price of the server is 2181.72 USD, so $p_{com} = 2181.72$ USD. As mentioned above, the Intel Xeon server is expected to run a Beacon node program at $E_{com} = 81$ W. By substituting $p_{com}$ and $E_{com}$ of the Intel Xeon server and equation (5,6,7,8,17) into equation (14) and (16), we obtain the upper limit of energy consumption and carbon footprint of Ethereum 2.0 are 0.3119 TWh and 0.1348 $MtCO_2$, respectively. In this scenario, the total number of Beacon nodes running on Ethereum 2.0 is 439507. Given the total number of full nodes on Ethereum 1.0 is less than 7000, the PoS-based Ethereum 2.0 is expected to be capable of achieving significantly higher level of decentralization with a much lower carbon footprint, comparing to PoW-based Ethereum 1.0.

**Projected Carbon Footprint of Blockchain Based on Logistic Model**

Predicting the carbon footprint of blockchain is particularly difficult as it involves order-of-magnitude changes, and includes many assumptions. Here, we propose another method of logistic model, as a supplement to the historic adoption model in the Main Text, to predict the carbon emissions from blockchain in the future.

$$P = \frac{K \times P_0}{P_0 + (K - P_0) \times \exp(-r_0 \times t)} \quad (18)$$

We try to estimate how many blockchain transactions would the world see today, if all financial transactions $K$ were covered by blockchain. The worldwide cashless transactions were $K = 779.1$ Billion times in 2020. The transactions on Bitcoin were 112,559,843 times in 2020, which has increased from 31,332 times in 2009. We used the data of Bitcoin transactions in 2009 to 2020 to fit the logistic curve to obtain the initial growth rate $r_0 = 0.219$ and the initial Bitcoin transaction number $P_0 = 9,714,478$. The Bitcoin transaction number $P$ of year $t$ can be derived by Eq. 18. Note that the data source has been listed in the Data Sheet 4.1. It is perfectly plausible that all cashless transactions could become supported by blockchain technology, and the carbon emission under this condition could be assumed as an upper limit from blockchain. We assume that the blockchain validation cost is proportional to the volume going through the system.

If Bitcoin and Ethereum follow the growth by a logistic curve, the projected carbon emission is shown in Fig. S2 (See details in Data: Sheet 4.1-4.6). If PoW-based Ethereum were to follow the logistic growth trend, the cumulative $CO_2$ emissions of this blockchain alone could warm the planet 0.29 to 0.47 °C by 2120, shown in Fig. S2b. Fig. S2d shows the total cumulative $CO_2$ emissions of PoW-based Ethereum and PoW-based Bitcoin crossing the 1.5 °C threshold within 50 years. Fig. S2e shows the carbon emission from PoS-based Ethereum and Bitcoin. The temperature increase from PoS-based Ethereum and Bitcoin can be reduced to 0.18 °C to 0.29 °C in comparison to the 6.86 °C to 11.18°C caused by PoW-based Ethereum and Bitcoin in 100 years.

The logistic model predicts a very early adoption of blockchain technology. The adoption is larger than 95% in 50 years. Whereas the method of historically based adoption in the Main Text is only at about 60% in 50 years. The difference is because blockchain has been experiencing a fast exponential growth at its early stage shown as Fig S1b, and the rapid growth rate will be constrained by the limitations on resources shortly based on the logistic model when $P_0 = K$. The exponential growth rate of blockchain in the early years is faster than the median curve of 36 widely adopted technology.

The logistic model provides another way to predict the carbon emission from blockchain compared to the model of adoption of broadly used technologies. However, accurately predicting the future carbon emission of blockchain is a challenge. The energy consumption of blockchain is not limited to financial transfer, such as digital assets that might have to be recorded and preserved in a ledger on blockchain. Furthermore, the energy consumption of blockchain is based on block mining difficulty, but future block mining difficulty is hard to predict. In addition, the carbon emission from per unit of electricity might decrease with the wide adoption of renewable energy in the future. Our model also does not consider this factor. Our study concentrates on the proposed PoS mechanism that could reduce carbon emissions of blockchain by more than 97%.

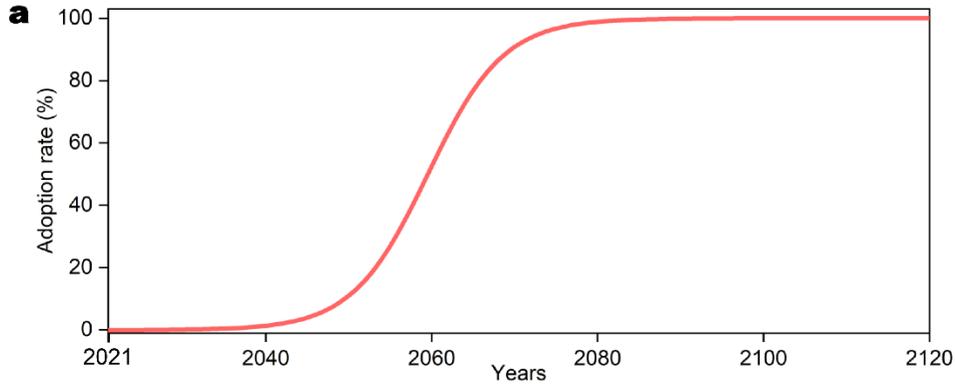
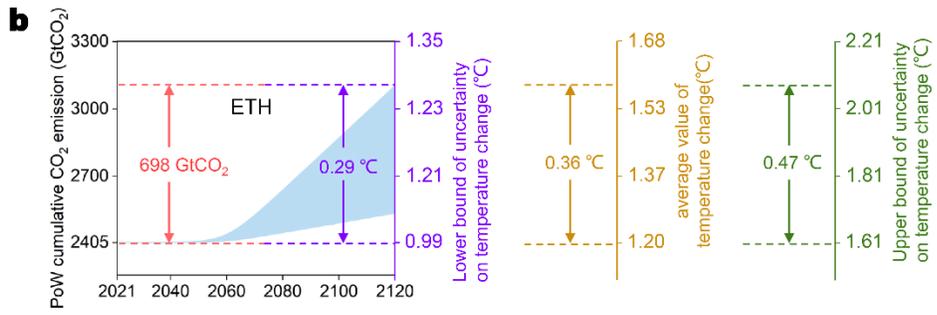
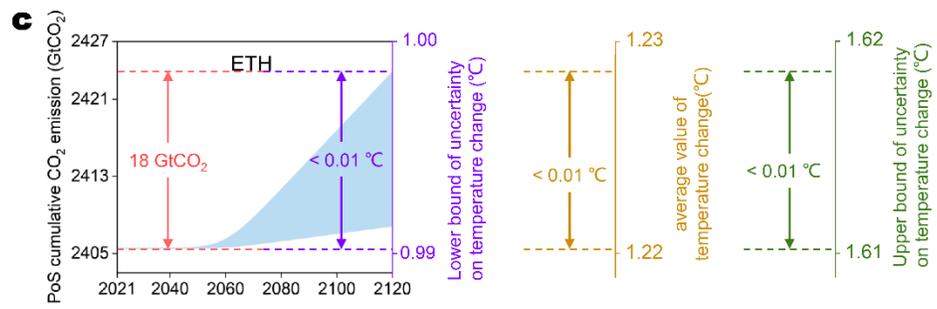
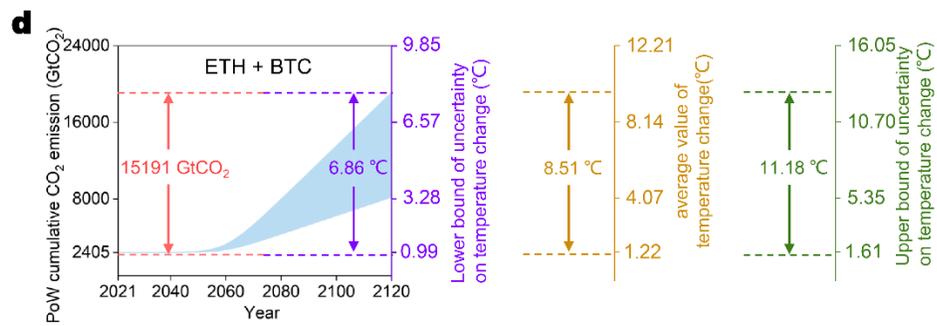
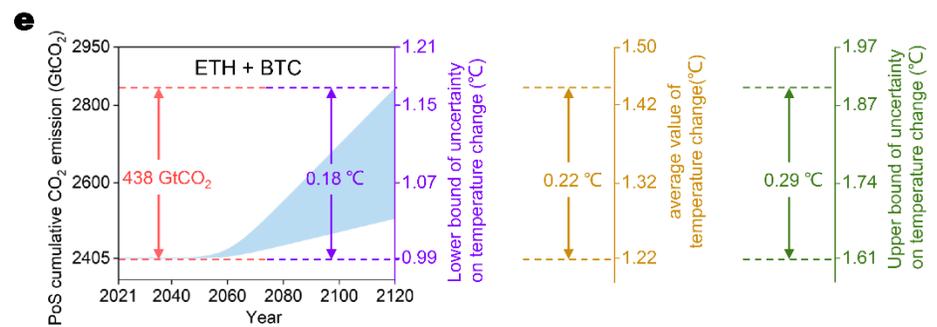

**Fig. S2** Projected Carbon Footprints of PoW and PoS-based Blockchain based on the logistic model. **Fig. S2a** shows trends in the adoption of blockchain technology based on the logistic model documented in Data: Sheet 4.1. **b** and **c** show the projected cumulative $CO_2$ emissions and the resulting temperature rise from PoW-based Ethereum 1.0 and PoS-based Ethereum 2.0, respectively, if they were to follow the growth rate of the logistic model in **a**. We conduct an uncertainty analysis on the relationship between carbon emissions and the increase in global temperature. Purple right Y axis shows lower bound of uncertainty by carbon emission on temperature change, orange right Y axis shows the average value on temperature change, and green right Y axis shows the upper bound on temperature change. **d** and **e** show the projected cumulative $CO_2$ emissions and the resulting temperature rise from PoW-based Ethereum 1.0 plus PoW-based Bitcoin, and PoS-based Ethereum 2.0 plus PoS-based Bitcoin, respectively, under the growth rate of the logistic model in **a**. The Data is shown in Data: Sheets 4.1–4.6.

**Current trend of Ethereum and Bitcoin usage.**

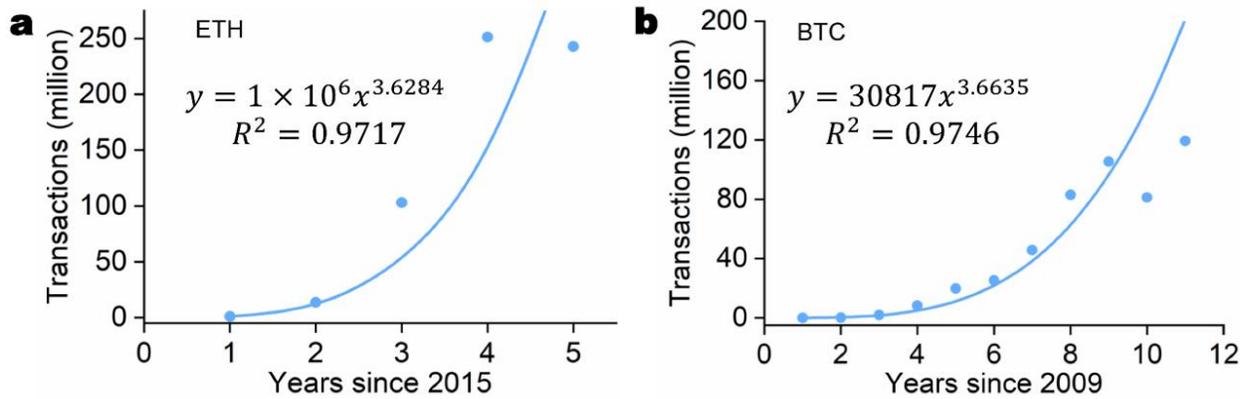

Fig. S1. Current trend of **a** Ethereum and **b** Bitcoin usage. The use of Ethereum blockchain for decentralized finance, health care, Internet-of-Things (IOT) networks, video games and so on has a fast growth since its introduction in 2015, that is currently best described with an exponential model. Bitcoin also follows an exponential growth. The Data is shown in Data sheet: 5.1